\documentclass[preprint2]{aastex}

\slugcomment{Submitted to ApJ}
\lefthead{Imanishi, Tsujimoto, \& Koyama}
\righthead {X-rays from Bona-fide and Candidate Brown Dwarfs in the
$\rho$ Oph cloud}

\begin{document}

\title{X-ray Detection from  Bona-fide and Candidate Brown Dwarfs in the
$\rho$ Ophiuchi Cloud with {\it Chandra}}

\author{Kensuke~Imanishi, Masahiro~Tsujimoto, and
Katsuji~Koyama}
\affil{Department of Physics, Graduate
School of Science, Kyoto University, Sakyo-ku, Kyoto, 606-8502, Japan}
\email{kensuke@cr.scphys.kyoto-u.ac.jp,
tsujimot@cr.scphys.kyoto-u.ac.jp, koyama@cr.scphys.kyoto-u.ac.jp}

\begin{abstract}

We present results of an X-ray search from bona-fide and candidate brown
dwarfs in the $\rho$ Ophiuchi cloud cores with the {\it Chandra} X-ray
Observatory.  The selected areas are two fields near the cloud center
and are observed with the ACIS-I array of a $17'\times 17'$ size and a
$\sim$100 ks exposure. Among 18 bona-fide and candidate brown dwarfs
listed by the infrared spectroscopy, we find X-ray emission from 7
sources above 99.9\% confidence level. Therefore $\sim$40\% of the
infrared-selected brown dwarfs in this cloud emit X-rays. For the
brightest 4 sources, the X-ray spectra are made and are fitted with a
thin-thermal plasma model of a temperature 1--2.5 keV. The X-rays are
also time variable with rapid flares from 2 of the brown dwarfs.
Assuming 2 keV temperature and using the empirical relation of $A_V$ vs. 
$N_{\rm H}$, we estimate the X-ray luminosity or its upper limit of the
other faint or non-X-ray sources. The X-ray luminosity ($L_{X}$) of the
X-ray-detected sources is in the range of 0.3--90$\times$10$^{28}$ ergs
s$^{-1}$, while the luminosity ratio of X-ray to bolometric
($L_{X}$/$L_{\rm bol}$) is 10$^{-3}$--10$^{-5}$, similar to those of
low-mass pre-main-sequence and dMe stars.  All these results suggest
that the X-ray origin of brown dwarfs is the same as low-mass stars;
strong magnetic activity at the stellar surface.
\end{abstract}


\section{INTRODUCTION}

Brown dwarfs are sub-stellar objects with mass well below the hydrogen
burning limit ($\sim$0.08 M$_{\odot}$), which fill up the gap between
stars and planets. With no stable nuclear burning, the energy source of
brown dwarfs is gravitational contraction, hence brown dwarfs become
cooler and less luminous with the increasing ages.  {\it ROSAT} detected
X-rays from several brown dwarfs and their candidates in star-forming
regions; Chamaeleon, $\rho$ Ophiuchi ($\rho$ Oph), and Taurus
\citep{Neuhauser1998, Neuhauser1999}. X-ray flares from field brown
dwarfs were also found \citep{Fleming2000, Rutledge2000}. These X-ray
features are similar to those of low-mass stars, hence the X-rays are
likely to be magnetic origin. However the standard ($\alpha$-$\omega$)
dynamo to produce the magnetic activity may not present, because brown
dwarfs are fully convective, hence there are no anchor points of the
magnetic field in the star interior \citep{Drake1996}. In order to
address the X-ray features and to study the emission mechanisms of brown
dwarfs, our current knowledge is still very poor, due mainly to the
limited sensitivity of the previous instruments.  This observational
constrain is largely relaxed by the {\it Chandra} X-ray Observatory of
wide band sensitivity (0.5--10.0 keV) coupled with unprecedented spatial
resolution of $\sim$0.5$''$ \citep{Weisskopf1996}.  We systematically
search and study X-rays from brown dwarfs in one of the nearest
star-forming regions, using the data of two deep ACIS-I exposures on the
$\rho$ Oph molecular cloud cores.

\section{OBSERVATIONS AND DATA REDUCTION}

Two {\it Chandra} observations were made on the central region of the
$\rho$ Oph cloud with the ACIS-I array consisting of four abutted X-ray
CCDs.  The first observation (here and after, obs.1) covered a
17\farcm4$\times$17\farcm4 area including cores B, C, E, and F, while
the second observation (obs.2) covered 
%
%
the center of core A \citep{Motte1998}. The level 2 data are
retrieved from the {\it Chandra} X-ray Center (CXC) archive, in which
the data degradation caused by the increase of charge transfer
inefficiency (CTI) in orbit is corrected.  X-ray events are selected
with the {\it ASCA} grades 0, 2, 3, 4, and 6. After the processing,
$\approx$100 ks effective exposure time is obtained from each
observation. The log of the observations is listed in Table
\ref{tab:obs}.

\placetable{tab:obs}

In these two fields, 18 late M dwarfs have been reported, based on the
water vapor absorption at $\lambda$ = 2.4--2.5 $\mu$m (Wilking, Greene,
\& Meyer 1999; Cushing, Tokunaga, \& Kobayashi 2000).  Among them, 8
sources have the upper limit of mass less than 0.08 M$_{\odot}$, which
we call ``bona-fide brown dwarfs'' (here and after, BDs). The other 10
sources have a mass in the transition region of 0.08 M$_{\odot}$, hence
called ``candidate brown dwarfs'' (CBDs). Their names and spectral types
are shown in Table \ref{tab:BDs}.

\placetable{tab:BDs}

\section{ANALYSIS AND RESULTS}

\subsection{Source Detection}

Using the \texttt{wavdetect}\footnote{see
http://asc.harvard.edu/udocs/docs/swdocs/detect/html/} command, we pick
up $\sim$100 X-ray sources from each field above the significance
criterion of 10$^{-7}$.  Infrared (IR) counterparts from the catalog of
\citet{Barsony1997} are searched and the position is cross-correlated to
that of the X-ray source.  Systematic coordinate offset of the {\it
Chandra} frame is then fine-tuned to fit the IR frame.  After the offset
correction, the relative position error (1$\sigma$) is 1\farcs2.  We
find X-ray counterparts from 5 IR positions in a 2$\sigma$ error radius
(2\farcs4) out of the 18 catalogued BDs and CBDs (GY 310, GY 31, GY 37,
GY 59, and GY 326). The X-ray positions and relative offsets from the IR
sources are given in Table \ref{tab:BDs}. The X-ray counts are extracted
from a circle of a half radius of the point-spread function (PSF) around
the X-ray position.

Since no apparent X-ray sources are found in a 2$\sigma$ error radius
(2\farcs4) of the other 13 catalogued BDs and CBDs, we define a circle
with a half radius of PSF around each IR position.  Then we manually
count the X-ray photons in the circle. We note that a rather small
source radius is selected so as to maximize the signal-to-noise (S/N)
ratio, particularly, for faint X-ray sources.  Nevertheless, as is
demonstrated in the 5 bright sources, the position error between IR and
X-rays is always smaller than the source radius, because both have
generally similar dependence on the source off-axis angle; both have the
smaller values for the sources with smaller off-axis angle.  Therefore
most of the X-ray photons for the relevant BDs and CBDs, if any, may
fall in the source circles.  The X-ray flux thus counted are given in
Table \ref{tab:BDs}.

The mean background counts are estimated from source-free regions of a
63-arcmin$^2$ and a 59-arcmin$^2$ area in the ACIS-I fields for obs.1
and obs.2, respectively.  The soft band (0.5--2.0 keV) background counts
(in units of 10$^{-2}$ counts arcsec$^{-2}$) are 2.3 and 2.2, for obs.1
and obs. 2, respectively, while those in the hard (2.0--9.0 keV) band
are about 3 times larger, 6.8 and 6.5 for obs.1 and obs. 2. The
background counts in each source area are given in Table 2. We then
separately calculate the confidence level ($CL$) of the X-ray detection
for the soft, hard and total (0.5--9.0 keV) bands. Based on the Poisson
statistics, the $CL$ is defined as;
\[ CL = \sum_{N'= 0}^{N_0-1} e^{-N_{bg}} \frac{N_{bg} ^{N'}}{N'!} \]
where $N_{0}$ and $N_{bg}$ are the detected and the background counts in
the source circle, respectively (see e.g., Gehrels 1986, and references
therein). We set the detection (or the upper limit) criterion that the
$CL$ should be larger than 0.999 (=3.3$\sigma$) in any of the 3 energy
band data.  In Table \ref{tab:BDs}, we show the maximum $CL$ value for
each source.  For sources located in both fields of view (GY 141, GY
163, and GY 202), we also estimate the $CL$ for the combined data as
well as the separate data. However, no higher $CL$ value is obtained
from all the sources.

As are summarized in Table \ref{tab:BDs}, bright X-rays are found from 3
CBDs (GY 31, GY 37, and GY 59) in addition to previously reported
sources, GY 310 and GY 326 (Imanishi, Tsuboi, \& Koyama 2001).
Detection of X-rays from a CBD, GY 84 is also highly significant.
Another BD, GY 141 in obs.1, show faint X-rays with $CL$ of $\geq$0.999
in the soft band. Using the log$N$--log$S$ relation by
\citet{Mushotzky2000}, we estimate the chance coincidence of a
background source to fall in the source circle to be 0.6--1 \%, where
the uncertainty mainly comes from the ambiguity of the amount of
interstellar absorption. Furthermore, background sources are most likely
AGNs and may have harder spectra than the possible counterpart of GY
141.  We thus conclude that the faint X-rays from this BD are real.

\subsection{X-ray Spectra and Luminosities}

\begin{figure}
 \figurenum{1}
 \epsscale{0.45}
 \plotone{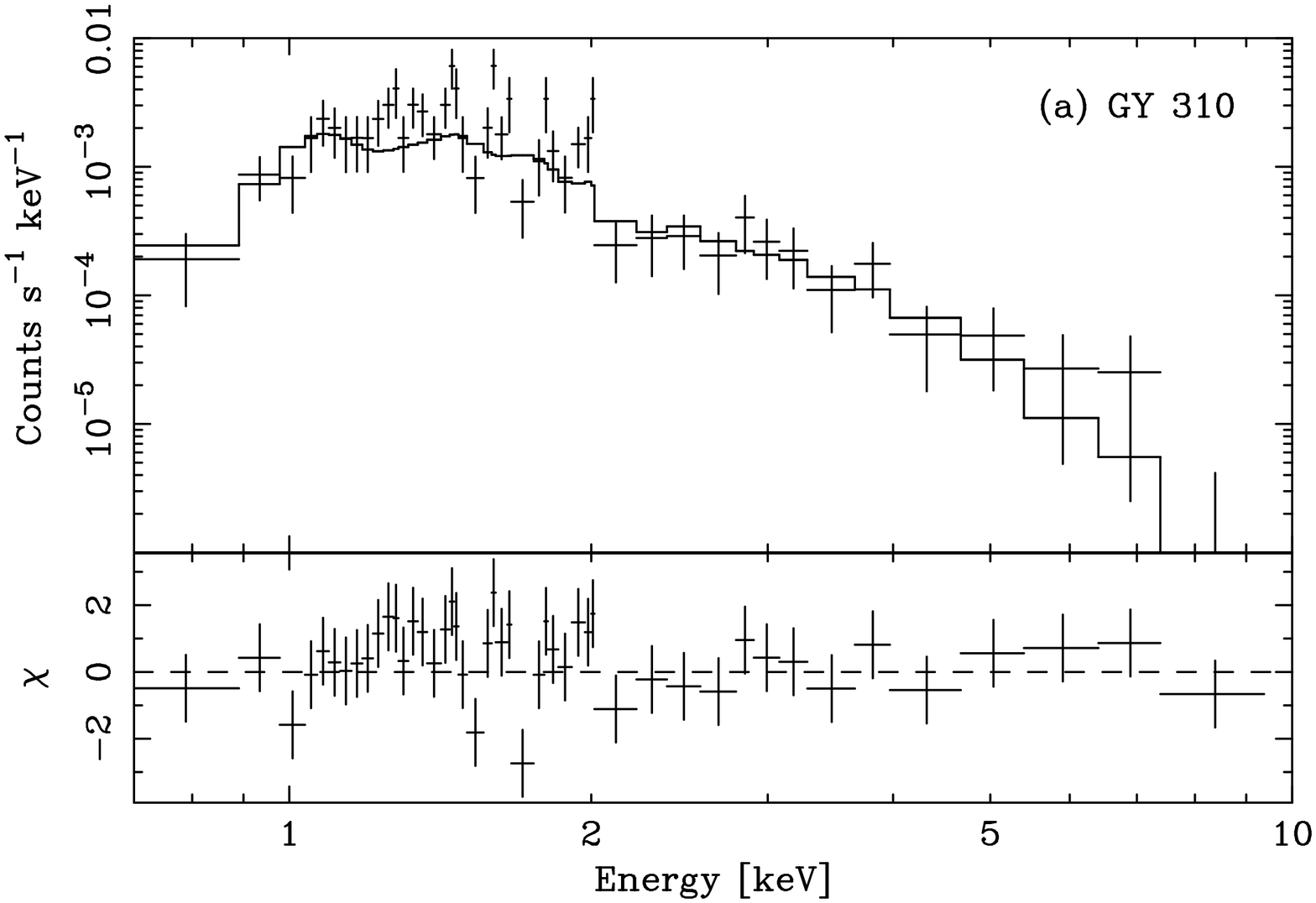}
 \plotone{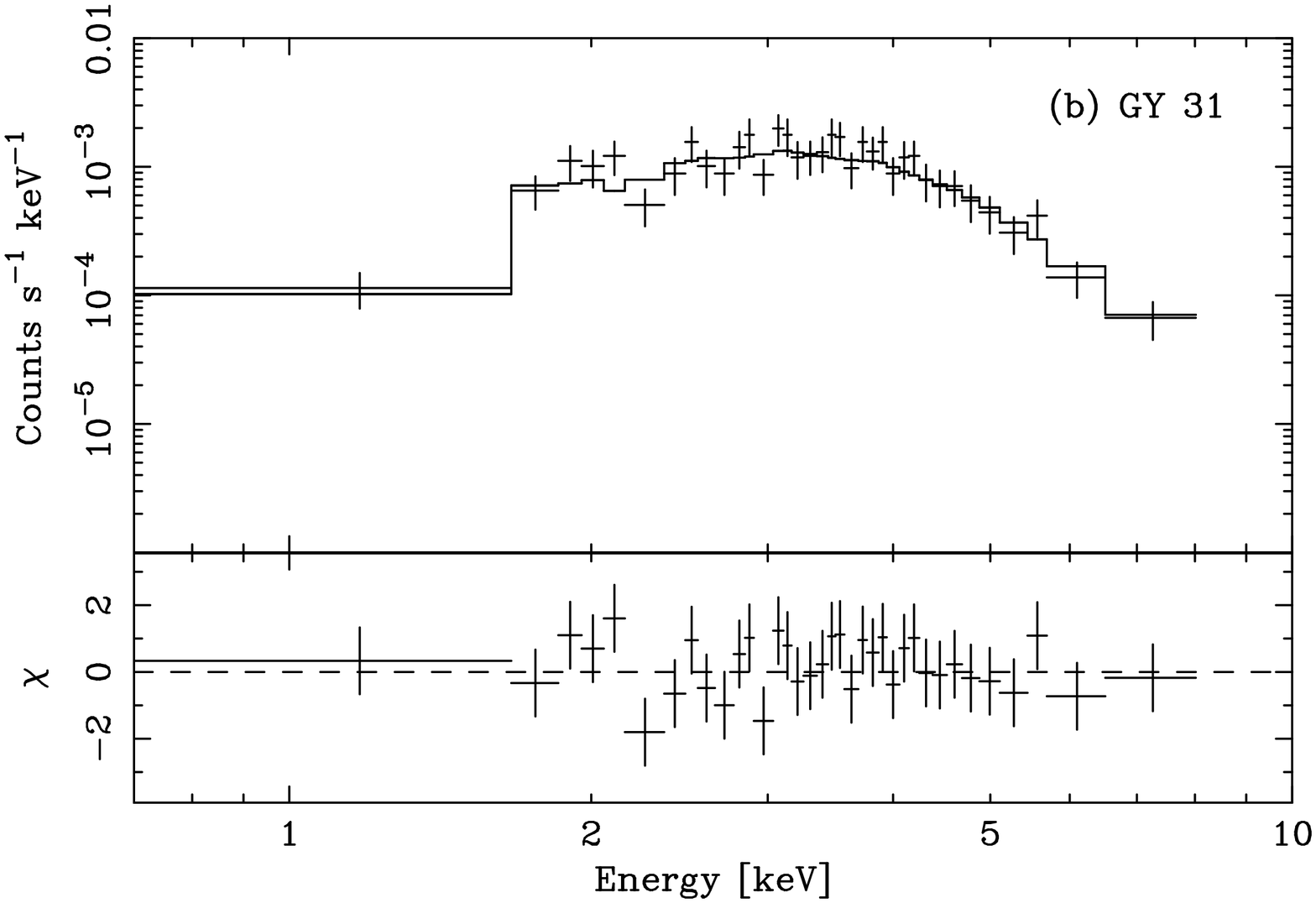}
 \plotone{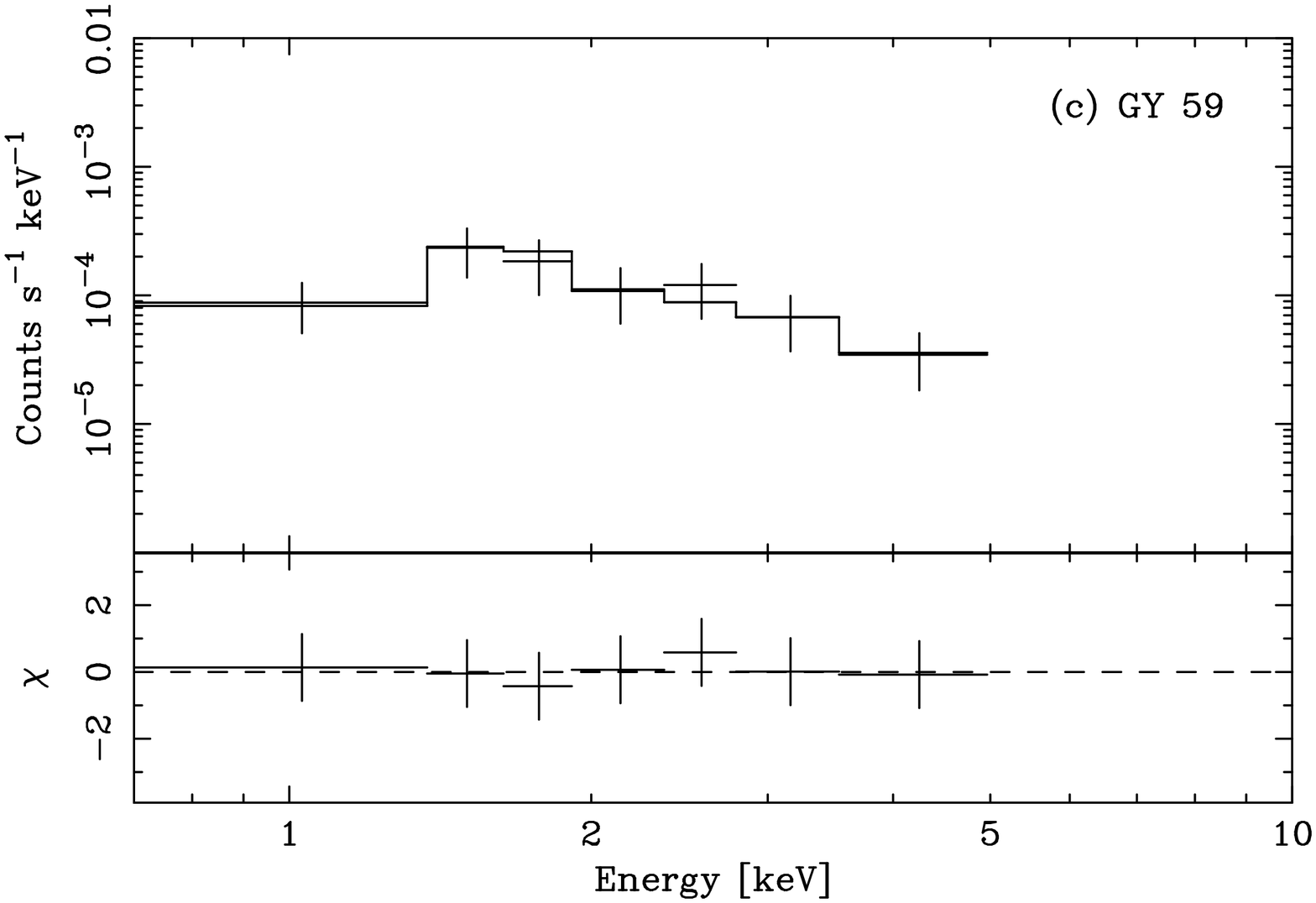}
 \plotone{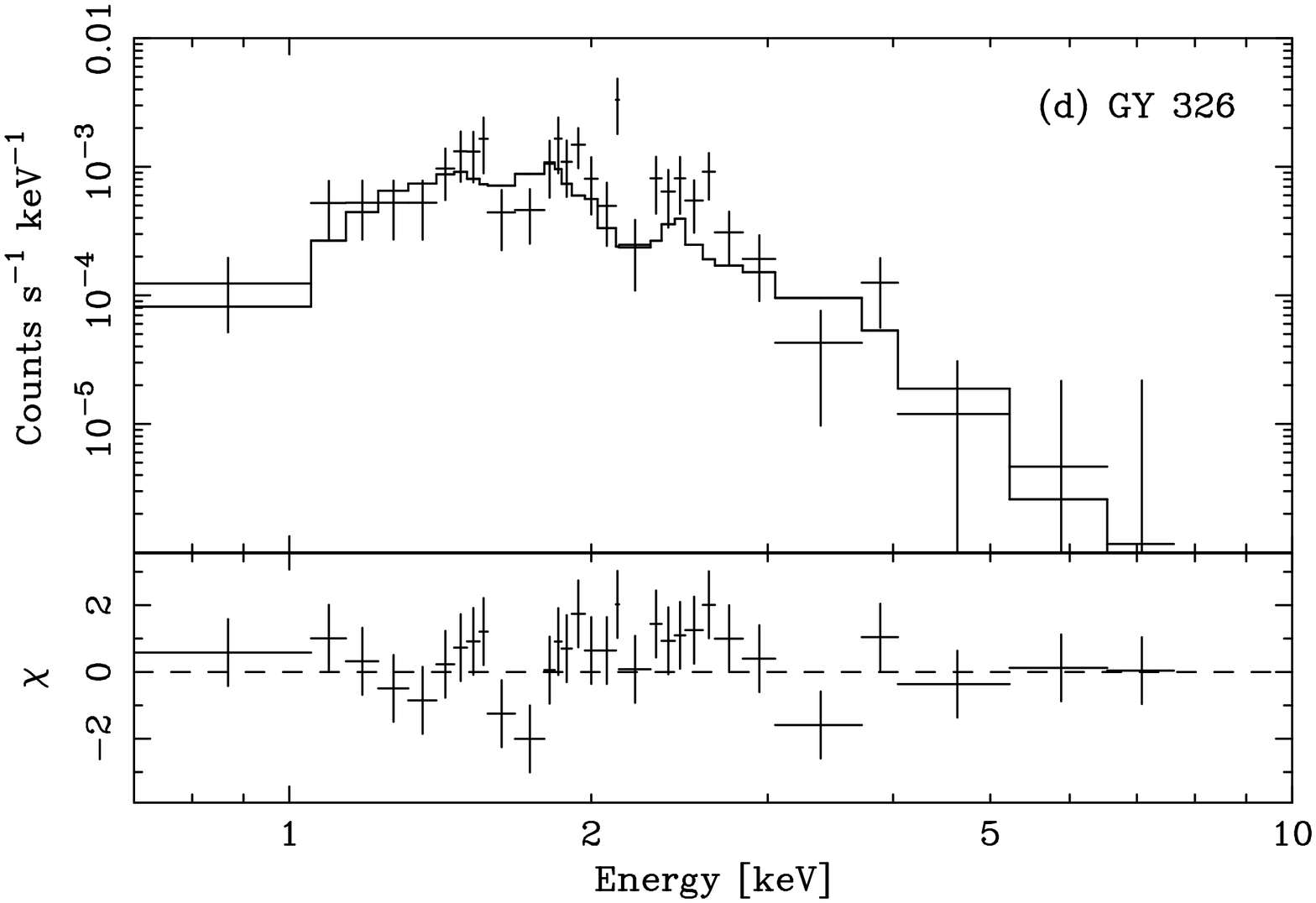}
 \caption[f1a.eps,f1b,ps,f1c.eps,f1d.eps]{Spectra of (a) GY 310, (b) GY
 31, (c) GY 59, and (d) GY 326. The upper panels show data points
 (crosses) and the best-fit model (solid line), while the lower panels
 are the data residuals from the best-fit model.  \label{fig:spec_bright}}
\end{figure}

For the brightest 4 sources ---, 1 BD (GY 310) and 3 CBDs (GY 31, GY 59,
and GY 326) ---, we analyze the time-averaged X-ray spectra. The
background spectra are extracted from the same regions used in \S3.1. We
%
%
then fit the spectra with a thin-thermal plasma (MEKAL: Mewe,
Gronenschild, \& van den Oord 1985) model, in which the metal abundance
is fixed to 0.3 solar based on the fitting results of other {\it
Chandra} X-ray sources in this region \citep{Imanishi2001}. The spectra
and the best-fit parameters are shown in Figure \ref{fig:spec_bright}
and Table \ref{tab:spec_bright}. The best-fit plasma temperature ($kT$)
is in the range of 0.9--2.5 keV.  Small differences in the parameters of
GY 310 and GY 326 from those in Table 1 of \citet{Imanishi2001}, which
is nevertheless unimportant, would be attributed to the more elaborated
data-reduction process of this paper (see \S2 in Imanishi et al.\ 2001).

For the faint or non-X-ray sources, the spectrum is assumed to be a
thin-thermal plasma of 2-keV temperature (the mean value for the bright
sources), while the absorption column is estimated from the empirical
relation, $N_{\rm H}$ = 1.59$\times$10$^{21}$ $A_V$ cm$^{-2}$
\citep{Imanishi2001}, where $A_V$ is the visual extinction derived from
IR photometry \citep{Wilking1999, Cushing2000}. We then calculate the
X-ray luminosities or those of the 99.9 \% upper limits in the soft band
(Table \ref{tab:BDs}).

\placefigure{fig:spec_bright}
\placetable{tab:spec_bright}

\subsection{Time Variability}

\begin{figure}
 \figurenum{2}
 \epsscale{0.45}
 \plotone{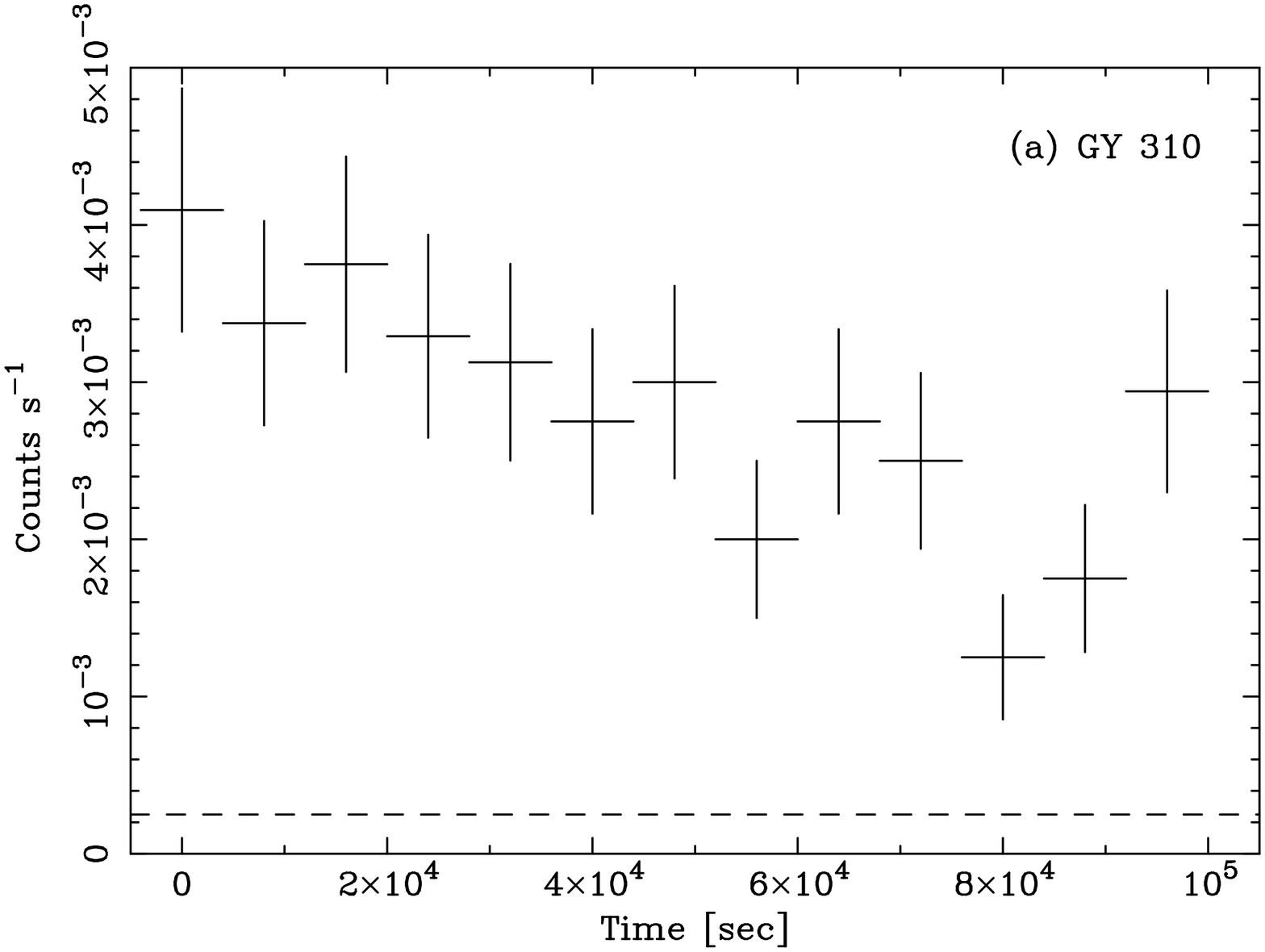}
 \plotone{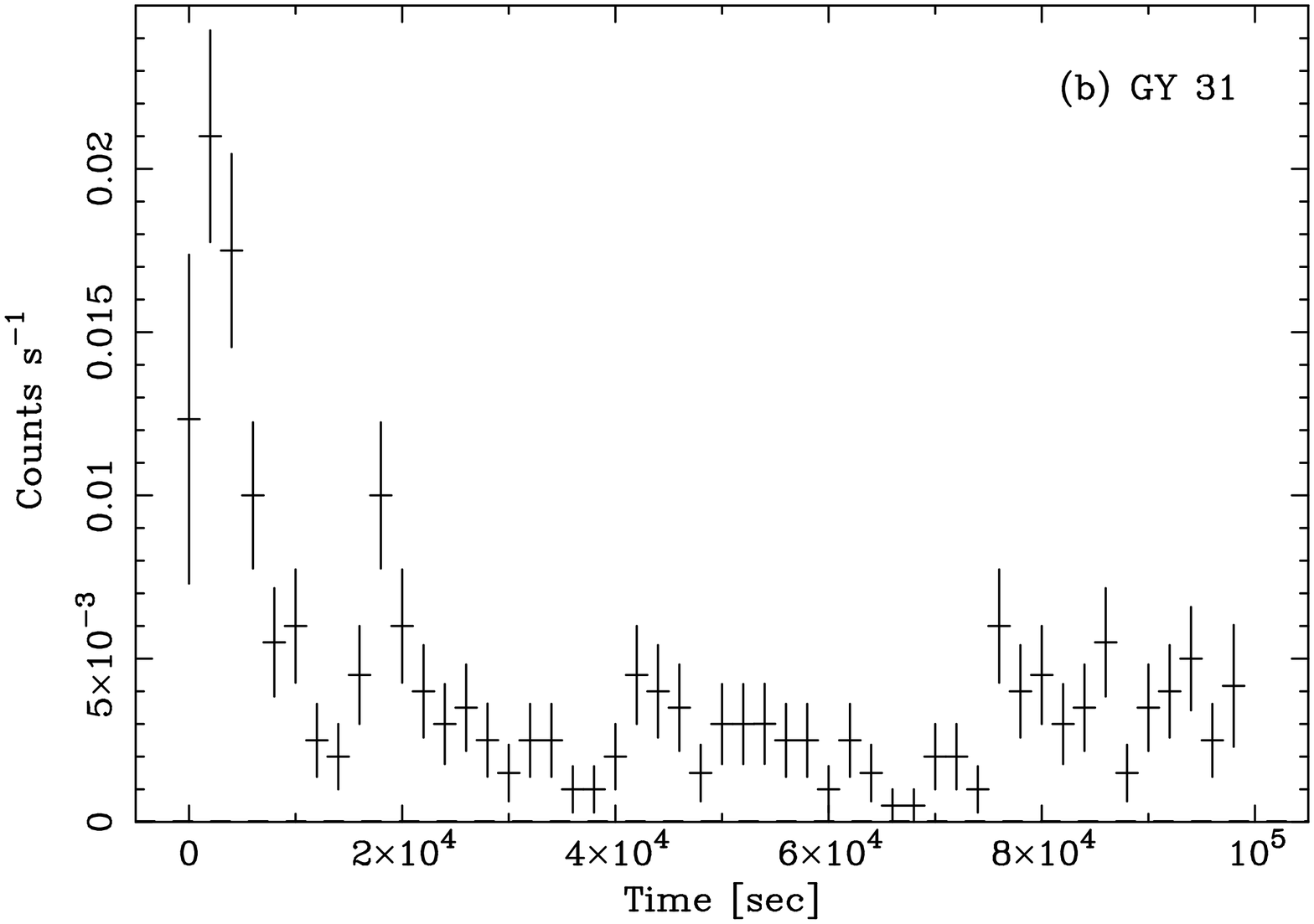}
 \plotone{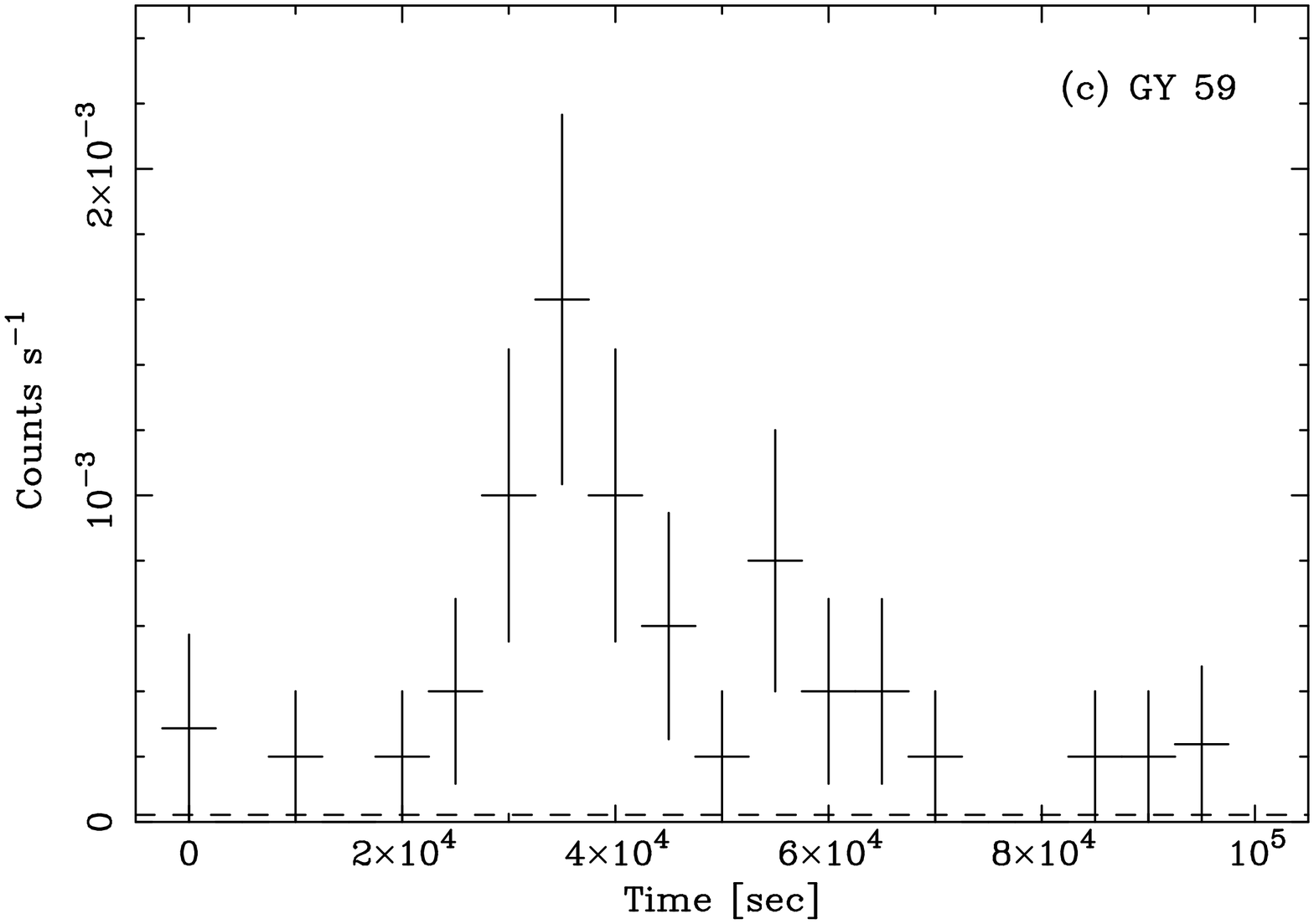}
 \plotone{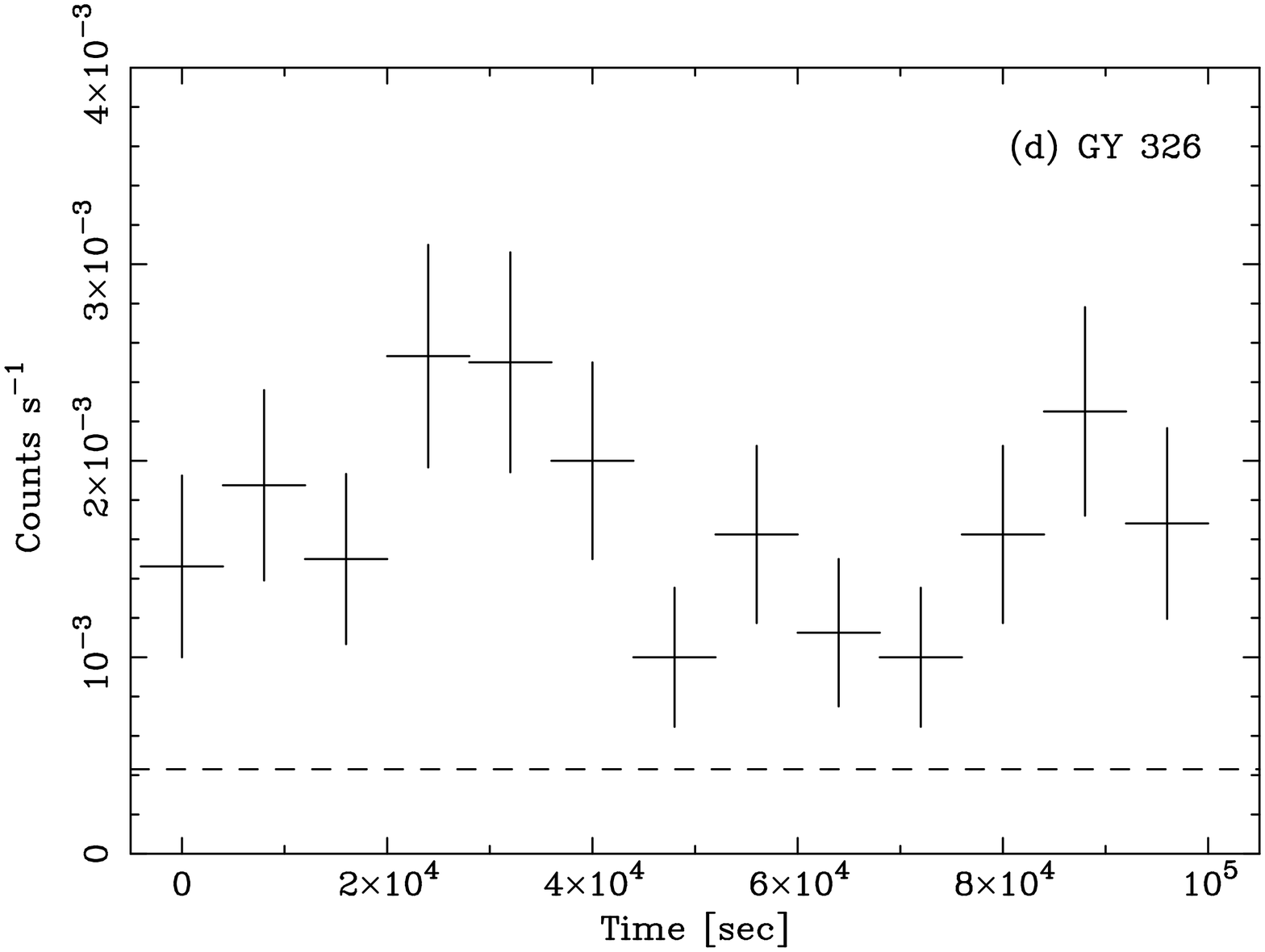}
 \caption[f2a.eps,f2b,ps,f2c.eps,f2d.eps]{Light curves of (a) GY 310, (b)
 GY 31, (c) GY 59, and (d) GY 326 in the 0.5--9.0 keV band. The
 background level is shown by the dashed line. For GY 31, the background
 line ($\sim$2$\times$10$^{-5}$ counts s$^{-1}$) overlaps with the bottom
 solid line.  \label{fig:lc_bright}}
\end{figure}

Figure \ref{fig:lc_bright} shows the light curves of (a) GY 310, (b) GY
31, (c) GY 59, and (d) GY 326 in 0.5--9.0 keV, where the background
levels are represented by the dashed lines.  For GY 31 and GY 59, clear
flare-like events, typical to those of low-mass main-sequence and
pre-main-sequence stars; a fast rise and slow decay, are seen. The
flare-decay timescale ($\sim$10$^4$ sec) is also the same as that of
low-mass stars.

GY 310 and GY 326, on the other hand, show no clear flare, but exhibit
aperiodic variability of flux change by a factor of $\sim$2 within 100
ks exposure. Since we see no large intensity variation in the background
flux, we conclude that the variability is highly significant.

In order to search for spectral change during the flare, we separately
make and fit the spectra of GY 31 in the flare and quiescent phases.  A
hint of spectral hardening during the flare is found, although it is not
significant from the statistical point of view; $kT$ is
2.5$^{+0.9}_{-0.7}$ keV at the first flare then decreases to
2.1$^{+0.6}_{-0.5}$ keV in a rather quiescent phases.

\placefigure{tab:lc_bright}

\section{DISCUSSION}

We detect X-ray emission from 2 out of 8 BDs and 5 out of 10 CBDs in the
$\rho$ Oph cloud cores. The X-ray detection rate is thus 25\% (BD) and
50\% (CBD). \citet{Neuhauser1999} claimed the X-ray detection of only 1
BD (GY202) with {\it ROSAT}/PSPC, in the same fields of the present
observations. Our present observations find no X-ray from GY202 with an
upper limit of 1.2$\times$10$^{28}$ ergs s$^{-1}$.  Instead, we find a
class I candidate WL 1 (No.13, in Imanishi et al.\ 2001) within the PSPC
error circle of GY202. This source emits X-rays with comparable
luminosity to that of GY202 reported with {\it ROSAT}/PSPC. Conversely,
\citet{Neuhauser1999} found no X-ray with an upper limit of
$\sim$10$^{28}$ ergs s$^{-1}$ from all the {\it Chandra} detected
sources.  They, however, estimated the X-ray luminosity assuming $kT = $
1 keV and $A_{\rm v}= $ a few mag (i.e., $N_{\rm H}$ = a few
$\times$10$^{21}$ cm$^{-2}$), which may be significantly lower than the
real case (see Table \ref{tab:spec_bright}). We hence re-estimate the
upper limit with the same assumption adopted in this paper (\S3.2), then
find that the {\it ROSAT} upper limit could be nearly one or two order
magnitude higher than those of the original paper. If we adopt these
newly estimated upper limits, we find no significant flux change from
the {\it ROSAT} to the {\it Chandra} era.
  
In Table \ref{tab:BDs}, we display the ratio of the X-ray ($L_X$) and
bolometric ($L_{\rm bol}$) luminosities. $L_{\rm bol}$ is roughly
proportional to the area of the photosphere, hence $L_{X}$/$L_{\rm bol}$
is a good indicator of 
%
%
coronal X-ray activity per unit surface
area. The $L_{X}$/$L_{\rm bol}$ value lies between 10$^{-3}$--10$^{-5}$,
which is similar to low-mass pre-main-sequence stars (e.g., Imanishi et
al.\ 2001) and dMe stars \citep{Giampapa1996}, but is significantly
higher than that of the quiescent solar corona; $\sim$5$\times$10$^{-7}$
\citep{Vaiana1978}. It should also be noted that most of the upper
limits of $L_{X}$/$L_{\rm bol}$ of X-ray non-detected BDs and CBDs are
scattered around 10$^{-3}$--10$^{-5}$, which are comparable with or
slightly lower than those of X-ray detected BDs and CBDs. This leads us
to suspect that these X-ray non-detected BDs and CBDs may emit X-rays
below but near at the current sensitivity limit of {\it Chandra}.

We also obtain the X-ray spectra from 1 BD and 3 CBDs for the first time
(\S3.2), which are fitted with a thin thermal plasma model of $\sim$2
keV temperature. Solar-like flares are also detected from 2 CBDs (GY 31
and GY 59, \S3.3). These X-ray features are similar to those of low-mass
stars. Together with the high $L_{X}$/$L_{\rm bol}$ value, we suggest
that X-rays from BDs and CBDs are attributable to magnetic activities
like low-mass stars. Detection of highly variable radio emission from GY
31, which probably arises from non-thermal gyro-synchrotron emission,
also indicates the existence of strong and time variable magnetic
activity (P. Andr\'e, private communication; see Neuh\"auser et
al. 1999).

\begin{figure}
 \figurenum{3}
 \epsscale{0.45}
 \plotone{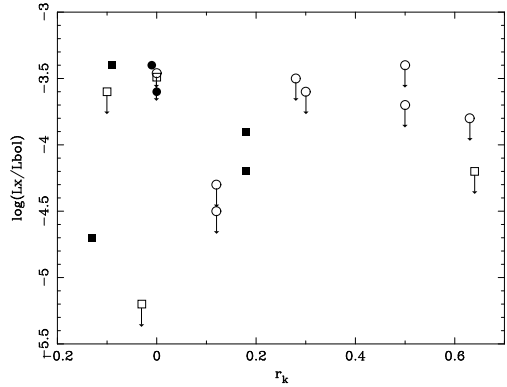}
 \caption[f3.eps]{Plots of the excess emission at $K$-band ($r_k$) and the
 luminosity ratio between X-ray and bolometric ($L_X$/$L_{\rm
 bol}$). Circles and squares represent BDs and CBDs, respectively. X-ray
 detection and non-detection are indicated by filled and open symbols,
 respectively.  Arrows indicate the upper limits.  GY 31 and GY 163 are
 not included in this figure because no $r_k$ value has been
 obtained. \label{fig:lx_rk}}
\end{figure}

One debatable issue is the mechanism of magnetic field amplification.
Brown dwarfs are fully convective, hence the standard dynamo
($\alpha$-$\omega$ dynamo) mechanism may not work. This situation is the
same as low-mass protostars.  As one possible scenario,
\citet{Montmerle2000} proposed an amplification of the magnetic field
between the star and the circumstellar disk by twisting through the
differential rotation. If BDs and CBDs emit X-rays with the same
mechanism, X-ray activity should vanish when the circumstellar disk
disappears as the star evolves.  Figure \ref{fig:lx_rk} shows the
relation between the excess emission at $K$-band and the $L_{X}/L_{\rm
bol}$ value. The excess emission is defined as $r_k$ = $F_{K_{\rm
ex}}$/$F_K$, where $F_{K_{\rm ex}}$ is the flux from excess
(circumstellar) emission and $F_K$ is the stellar flux at $K$-band
\citep{Wilking1999, Cushing2000}, hence $r_k >$ 0 indicates the presence
of the circumstellar disk. X-ray detected BDs and CBDs are uniformly
distributed over two regions of $r_k>$ 0 and $r_k\leq$ 0; we see no hint
that X-ray activity decreases as the circumstellar disk disappears
($r_k\leq$ 0). We therefore conclude that the existence of the disk is not
the key parameter in X-ray emission of BDs and CBDs.  Another type of
dynamo, such as turbulent-driven ($\alpha^2$) dynamo \citep{Durney1993},
is conceivable.

\placefigure{fig:lx_rk}

\acknowledgments
%
%
The authors express their thanks to Dr. Thomas Preibisch for critical
refereeing and useful comments. The authors also acknowledge Dr.\ Yohko
Tsuboi and Jun Yokogawa for useful discussions and comments. K.I.\ and
M.T.\ are financially supported by JSPS Research Fellowship for Young
Scientists.

\onecolumn

\clearpage

\begin{deluxetable}{lclccc}
 \tabletypesize{\scriptsize}
 \tablewidth{0pt}
 \tablenum{1}
 \tablecaption{Log of the {\it Chandra} ACIS-I Observations on the
 $\rho$ Oph Cloud Cores \label{tab:obs}}
 \tablehead{\colhead{Obs.ID} & \colhead{Sequence ID} & \colhead{Date} &
 \colhead{R.A.\tablenotemark{a}} & \colhead{Dec.\tablenotemark{a}} &
 \colhead{Exposure} \\
 & & & (J2000) & (J2000) & (ks)}
 \startdata
 1 & 200060 & 2000 Apr 13--14 & 16$^{\rm h}$27$^{\rm m}$18\fs1 & $-$24$^{\circ}$34$'$21\farcs9 & 100.6 \\
 2 & 200062 & 2000 May 15--17 & 16$^{\rm h}$26$^{\rm m}$35\fs3 & $-$24$^{\circ}$23$'$12\farcs9 & \phn96.4 \\
 \enddata
 \tablenotetext{a}{The position of the telescope optical axis.}
\end{deluxetable}

\begin{deluxetable}{lllcccrrrrcrrl}
 \rotate
 \tabletypesize{\scriptsize}
 \tablewidth{0pt}
 \tablenum{2}
 \tablecaption{Bona-fide and Candidate Brown Dwarfs in the $\rho$ Oph
 Cloud Cores \label{tab:BDs}}
 \tablehead{\colhead{Name} & \colhead{Sp.\ type\tablenotemark{a}} &
 \colhead{ID\tablenotemark{b}} & \colhead{R.A.\tablenotemark{c}} &
 \colhead{Dec.\tablenotemark{c}} & \colhead{offset\tablenotemark{d}} &
 \colhead{PSF\tablenotemark{e}} & \multicolumn{3}{c}{X-ray
 counts\tablenotemark{f}} & \colhead{1 $-$ $CL$\tablenotemark{g}} &
 \colhead{$L_X$\tablenotemark{h}} & \colhead{log($L_X$/$L_{\rm bol}$)} &
 \colhead{Ref.}
 \\
 & & & (J2000) & (J2000) & ($''$) & ($''$) & Soft & Hard & BGD & &
 (10$^{28}$ergs s$^{-1}$) & & }
 \startdata
 \sidehead{Brown Dwarf (BD)}
 CRBR 14 & M7.5 & 2 & \nodata            & \nodata        & \nodata & 5.7  & 3   & 2   & 0.5 & ND (S)  & $<$0.55 & $<-$4.3 & 1 \\
 GY 10   & M8.5 & 2 & \nodata            & \nodata        & \nodata & 3.2  & 2   & 0   & 0.2 & ND (S)  & $<$0.48 & $<-$4.5  & 1 \\
 GY 11   & M6.5 & 2 & \nodata            & \nodata        & \nodata & 3.3  & 1   & 1   & 0.2 & ND (T)  & $<$0.12 & $<-$3.8 & 1, 2 \\
 CRBR 31 & M6.7 & 2 & \nodata            & \nodata        & \nodata & 7.1  & 0   & 2   & 0.9 & ND (H)  & $<$0.56 & $<-$3.6 & 2 \\
 GY 64   & M8   & 2 & \nodata            & \nodata        & \nodata & 3.7  & 1   & 0   & 0.2 & ND (S)  & $<$0.93 & $<-$3.5 & 1, 2 \\
 GY 141  & M8   & 1 & \nodata            & \nodata        & \nodata & 9.0  & 7   & 1   & 1.5 
 & \phm{>} 8.0$\times$10$^{-4\phn}$ (S)& 0.25    & $-$3.6  & 2 \\
         &      & 2 & \nodata            & \nodata        & \nodata & 22.3 & 11  & 23  & 8.6 & ND (S)  & $<$0.33 & $<-$3.5 &   \\
 GY 202  & M7   & 1 & \nodata            & \nodata        & \nodata & 9.4  & 2   & 4   & 1.6 & ND (S)  & $<$1.2\phn  & $<-$3.7 & 1 \\
         &      & 2 & \nodata            & \nodata        & \nodata & 16.8 & 3   & 9   & 4.9 & ND (S)  & $<$2.0\phn  & $<-$3.4 &   \\
 GY 310  & M8.5 & 1 & 27$^{\rm m}$38\fs6 & 38$'$39\farcs1 & 1.3     & 9.2  & 182 & 59  & 1.5 
 & $<$1.0$\times$10$^{-15}$ (T) & 9.3\phn & $-$3.4  & 1 \\
 \sidehead{Candidate Brown Dwarf (CBD)}
 CRBR 15 & M5   & 2 & \nodata            & \nodata        & \nodata & 4.2  & 2   & 1   & 0.3 & ND (S)  & $<$1.1\phn & $<-$4.2 & 1 \\
 GY 5    & M7   & 2 & \nodata            & \nodata        & \nodata & 4.7  & 2   & 1   & 0.4 & ND (S)  & $<$0.16 & $<-$5.2 & 1 \\
 GY 31   & M5.5 & 2 & 26$^{\rm m}$25\fs2 & 23$'$25\farcs7 & 0.2     & 2.5  & 41  & 350 & 0.1 
 & $<$1.0$\times$10$^{-15}$ (T) & 93\phd\phn\phn & $-$3.5  & 1 \\
 GY 37   & M6   & 2 & 26$^{\rm m}$27\fs8 & 26$'$43\farcs9 & 0.8     & 4.2  & 7   & 5   & 0.3 
 & \phm{>} 3.7$\times$10$^{-8\phn}$ (S) & 0.59 & $-$4.2  & 1 \\
 GY 59   & M6   & 2 & 26$^{\rm m}$31\fs4 & 25$'$32\farcs2 & 0.7     & 2.6  & 17  & 22  & 0.1 
 & $<$1.0$\times$10$^{-15}$ (T) & 2.6\phn & $-$3.9  & 1 \\
 GY 84   & M6   & 2 & \nodata            & \nodata        & \nodata & 1.9  & 2   & 2   & 0.1 
 & \phm{>} 1.3$\times$10$^{-4\phn}$ (T)  & 0.85     & $-$4.7  & 1 \\
 GY 163  & M2.5 & 1 & \nodata            & \nodata        & \nodata & 12.4 & 5   & 12  & 2.8 & ND (T)  & $<$12\phd\phn\phn   & $<-$3.8 & 1 \\
         &      & 2 & \nodata            & \nodata        & \nodata & 11.9 & 3   & 7   & 2.4 & ND (S)  & $<$10\phd\phn\phn   & $<-$3.8 &   \\
 GY 218  & M4.0 & 1 & \nodata            & \nodata        & \nodata & 7.2  & 2   & 3   & 0.9 & ND (S)  & $<$0.45 & $<-$3.6 & 2 \\
 CRBR 67 & M4.4 & 1 & \nodata            & \nodata        & \nodata & 3.2  & 0   & 0   & 0.2 & ND (--) & $<$0.31 &  $<-$3.5 & 2 \\
 GY 326  & M4   & 1 & 27$^{\rm m}$42\fs8 & 38$'$51\farcs1 & 1.5     & 11.4 & 67  & 65  & 2.3 
 & $<$1.0$\times$10$^{-15}$ (T) & 28\phd\phn\phn      & $-$3.4  & 1 \\
 \enddata 
 \tablenotetext{a}{Spectral type determined from the water vapor
 absorption.}
 \tablenotetext{b}{Observation identification (see Table
 \ref{tab:obs}).}
 \tablenotetext{c}{Position of the identified X-ray source which is
 picked up with the \texttt{wavdetect} command (see \S3.1). Right
 ascension and declination for all sources are at 16$^{\rm h}$ and
 $-$24$^{\circ}$.}
 \tablenotetext{d}{Offset between the IR and the X-ray sources.}
 \tablenotetext{e}{Radius of the point-spread function.}
 \tablenotetext{f}{X-ray counts in the soft (0.5--2.0 keV) and hard
 (2.0--9.0 keV) bands, and predicted background counts in the soft band
 within the source circle. Background counts in the hard band is $\sim$3
 times larger than those in the soft band.}
 \tablenotetext{g}{$CL$ is the confidence level of the source detection
 (see \S3.1). Parentheses indicate the energy band in which the maximum
 $CL$ value is obtained (S: 0.5--2.0 keV, H: 2.0--9.0 keV, T: 0.5--9.0
 keV). ND represents non X-ray detection with $CL <$ 0.999 (i.e., 1$-CL
 >$ 10$^{-3}$).}
 \tablenotetext{h}{Absorption-corrected luminosity in 0.5--9.0 keV. The
 %
 %
 distance to each source is assumed to be 145 pc \citep{deZeeuw1999}.}
 \tablerefs{(1):~\citet{Wilking1999}; (2):~\citet{Cushing2000}}
\end{deluxetable}

\begin{deluxetable}{lcccc}
 \tabletypesize{\scriptsize}
 \tablewidth{0pt}
 \tablenum{3}
 \tablecaption{Best-Fit Parameters of the Brightest 4 Sources
\label{tab:spec_bright}}
 \tablehead{\colhead{Name} & \colhead{$kT$\tablenotemark{a}} &
 \colhead{log($EM$)\tablenotemark{a}} &
 \colhead{$N_{\rm H}$\tablenotemark{a}} & \colhead{$\chi^2$
 /$d.o.f$} \\
 & (keV) & (cm$^{-3}$) & (10$^{22}$cm$^{-2}$) & }
 \startdata
 GY 310 & 1.7(0.9--2.2) & 51.6(51.5--52.1) & 0.8(0.5--1.5) & 57.8/44 \\
 GY 31  & 2.2(1.7--2.9) & 52.6(52.4--52.8) & 5.9(5.1--7.2) & 25.1/33 \\
 GY 59  & 2.5($>$1.0)   & 51.0(50.6--51.7) & 1.4(0.5--3.0) & 0.556/4 \\
 GY 326 & 0.9(0.7--1.2) & 52.1(51.7--52.4) & 2.3(1.6--3.0) & 34.8/28 \\
 \enddata 
 \tablenotetext{a}{Parentheses indicate the 90\% confidence limits.}
\end{deluxetable}

\end{document}